\documentclass[aps,prd,twocolumn,showpacs]{revtex4}

\usepackage{amsmath,amssymb,graphicx}

\newcommand{\be}{\begin{equation}}
\newcommand{\ee}{\end{equation}}
\newcommand{\bea}{\begin{eqnarray}}
\newcommand{\eea}{\end{eqnarray}}

\begin{document}
\title{Black holes with non-minimal derivative coupling}

\author{Massimiliano Rinaldi}
\email{mrinaldi@fundp.ac.be}
\affiliation{Namur Center for Complex systems (naXys),\\ University of Namur, Belgium}

\date{\today}

\begin{abstract}
\noindent We study the gravitational field equations in the presence of a coupling between the derivative of a massless scalar field and the Einstein tensor. This configuration is motivated by Galileon gravity as it preserves shift invariance in the scalar sector. We analytically obtain solutions with static and spherically symmetric geometry, which also include black holes with a single regular horizon. We examine the thermodynamical properties of these solutions, and we reveal the non-perturbative nature of the coupling constant. We also find a phase transition, similar to the one described by Hawking and Page, which occurs at a critical temperature determined by both the black hole mass and by the strength of the coupling.
\end{abstract}


\maketitle

\section{Introduction}

\noindent In recent years, many extensions of general relativity have been considered in the attempt to explain dark energy. In particular, a lot of work has been devoted to the most general tensor-scalar action that generates equations of motion with second-order derivatives, discovered many years ago by Horndeski \cite{horn}. In fact, this action shows interesting self-tuning cosmological solutions \cite{fabfour}, and includes Galileon gravity \cite{galileon} and massive gravity \cite{massivegrav}. Moreover, it can be obtained by compactifying a suitable truncation of higher-dimensional Lovelock gravity \cite{christos}. Among the many interesting features of  Horndeski gravity there is the coupling between the derivative of a scalar field and the Einstein tensor.  On cosmological backgrounds, this term leads to accelerated expansion without the need of any scalar potential, as noted for the first time in \cite{amendola}. A similar accelerating effect was also observed in the case when the coupling involves the Ricci tensor alone \cite{Rcoupling}.  These features attracted much interest in both inflationary \cite{Sushkov,germani,namur} and late-time cosmology \cite{saridakis}, but little attention has been paid to local and  stationary solutions so far. 

In this paper, we wish to partially fill this gap by considering static geometries with spherical symmetry in order to find exact black hole solutions  with regular horizons. Approximate solutions in the presence of an electromagnetic field were found in \cite{greeks}. In a recent paper, it has been shown that a no-hair theorem is at work for Galileon gravity. In other words, there cannot be asymptotically flat black hole solutions unless the scalar field is globally constant \cite{nicolis}. Based on this result, Nicolis and Hui have proposed an interesting test to constrain the parameters of Galileon gravity \cite{nicolisexp}.  To evade the no-hair theorem we need to relax at least one hypothesis and the most natural one is asymptotic flatness, an idea supported by the existence of asymptotically de Sitter solutions. 

This strategy turns out to be successful and we are able to find analytic solutions with spherical symmetry and a regular horizon to the equations of motion. The paper is organized as follows. In Sec.\ 2 we obtain the equations of motion and we show that spherically symmetric solutions with a regular horizon exist in an analytic form.  In Sec.\ 3 we focus on the the thermodynamical properties of the black hole and the r™le of the coupling parameter. In Sec.\ 4 we briefly discuss other solutions in different areas of the parameter space and we conclude in Sec.\ 5 with some remarks.

\section{Equations of motion and analytic solutions}

To begin with, let us consider the Lagrangian 
\bea\label{lagr}
L= {m_p^{2}\over 2}R-{1\over 2}\left(g^{\mu\nu}-{z\over m_p^{2}}G^{\mu\nu}\right)\partial_{\mu}\varphi\partial_{\nu}\varphi,
\eea
where $m_{p}$ is the Planck mass, $z$ a real number, $G_{\mu\nu}$ the Einstein tensor,  $\varphi$ a scalar field, and $g_{\mu\nu}$ is the metric, chosen with mostly plus signature. The absence of scalar potential allows for the shift symmetry $\varphi\rightarrow \varphi+$const, which is the relevant Galileon symmetry that survives in curved space \cite{galileon}. For this reason we will refer to $\varphi$ as to the Galileon field in the following. In the usual Galileon terminology, the three terms in $L$ are representative of $L_{2}$, $L_{4}$ and $L_{5}$, see e.g \cite{steer}. The term involving the Einstein tensor also appears in the context of massive gravity, where the parameter $z$ is related to the inverse of the graviton mass \cite{massivegrav}. Our goal is to obtain static solutions with spherical symmetry, so the metric ansatz is
\bea
ds^{2}=-F(r)dt^{2}+G(r)dr^{2}+\rho^{2}(r)(d\theta^{2}+\sin^{2}\!\theta d\phi^{2}).
\eea
To obtain the equations of motion, we find more convenient to express the  Lagrangian in terms of the metric components and then vary the action $S=\int d^{4}x \sqrt{g}L$  with respect to the four fields $\varphi,\, F,\, G,\, \rho$. Once the equations of motion are known, one can set $\rho=r$ and find
\begin{align}
& r{F'\over F}={G-1}+{m_p^{2}r^{2}G\over z}+{Km_p^{2}G^{2}\over z \psi\sqrt{FG}}, \label{current}\\
& r{F'\over F}={2m_p^{4}G(G-1)+z\psi^{2}(G-3)+m_p^{2}r^{2}G\psi^{2}\over (3z \psi^{2}+2m_p^{4}G)},\\
& {r\over 2}\left({F'\over F}-{G'\over G}\right)=  {2m_{p}G(G-1)-2z\psi^{2}-zr(\psi^2)'\over (3z \psi^{2}+2m_p^{4}G) },  \label{bern}
\end{align}
where $K$ is an integration constant and $\psi\equiv\varphi'$. There is also a fourth, redundant equation  whose form is unimportant. We immediately note that  $\psi=0$  implies $K=0$ and the resulting metric turns out to be the Schwarzschild one. If $z=0$ (and $\psi\neq 0$), one finds the Just solution, which is known to be singular both at the origin and at the horizon, see e.g. \cite{damour}. Finally,  we note that the first equation includes the term $\sqrt{FG}$. Therefore, for $K\neq 0$, $G$ and $F$ must have the same sign for all $r$ for the metric components to be real-valued.

When $K=0$ and $z\neq 0$, we can analytically find exact solutions to the system (\ref{bern}). Their form crucially depends on the sign of $z$ and in this letter we mainly focus on the case $z>0$ as it is the most interesting. We will comment on the case with negative $z$ at the end. 
The solution to the system reads
\bea\label{gtt}
F(r)&=&{3\over 4}+{r^{2}\over l^{2}}-{2M\over m_{p}^{2}r}+{\sqrt{z}\over 4m_{p}r}\arctan\left(m_{p}r\over \sqrt{z}\right),\\
G(r)&=&{(m_{p}^{2}r^{2}+2z)^{2}\over 4(m_{p}^{2}r^{2}+z)^{2}F(r)},\\
\psi^{2}(r)&=&-{m_{p}^{6}r^{2}(m_{p}^{2}r^{2}+2z)^{2}\over 4z (m_{p}^{2}r^{2}+z)^{3}F(r)},\label{psi}
\eea
where we defined $l^{2}=12z/m_{p}^{2}$ and $M$ is a constant of integration that will play the role of a mass. There is a second constant that multiplies $F(r)$ and that can be absorbed into a redefinition of the time coordinate. We immediately note that the function $F$ is very similar to the $g_{tt}$ component of a Schwarzschild Anti-de Sitter (SAdS) black hole with spherical horizon \cite{hp,danny}. The analysis of the curvature invariants reveals that these are all finite for  $r>0$, and, in particular,  at $r=r_{h}$, namely at the  zero of the function $F(r)$, which is unique if $M>0$.  On the opposite, the Ricci scalar diverges at $r=0$ confirming that there is a physical singularity at the origin. Thus, the solution above describes a genuine black hole with one regular horizon located at $r=r_{h}$ if $M>0$. In contrast, when $M=0$ the metric is non-singular for all  $r\geq 0$ and this vacuum solution will be very important for the analysis of the thermodynamical properties \footnote{The curvature invariants were computed with Maple implemented with the GRTensor II package, see http://grtensor.phy.queensu.ca}. 

From Eq.\ (\ref{psi}) we see that $\psi$ vanishes only when $z\rightarrow \infty$. In fact, in this limit we recover the Schwarzschild solution with  $F=G^{-1}=1-2M/(m_{p}^{2}r)$ and $R_{\mu\nu}=0$.  This is also due to the fact that the Galileon term is constant on shell as $(g^{rr}-zG^{rr}/m_{p}^{2})\psi^{2}=-m_{p}^{4}/z$ for all $r$. All these elements indicate that $z$ is a non-perturbative parameter when we regard the Lagrangian (\ref{lagr}) as a theory of modified gravity. Indeed, the deviation from general relativity vanishes when $z$ diverges and the Galileon field is strongly coupled. Also, the parameter $z$ clearly interpolates between the flat black hole solution and the SAdS one as $1/z$ essentially plays the role of  an effective negative cosmological constant. 

\section{Thermodynamical properties}

The similarity between the geometry of our solution and the one of the SAdS black hole  suggests standard techniques to study the thermodynamical properties \cite{hp}. First of all, the inverse temperature $\beta$ is determined by the periodicity of the Euclidean metric obtained by the analytic continuation $t\rightarrow -i\tau$, that is \cite{peet}
\bea
\beta={4\pi\sqrt{g_{\tau\tau}g_{rr}}\over g'_{\tau\tau}}\Bigg|_{r=r_{h}}={8\,\pi z r_{h}\over (m_{p}^{2}r_{h}^{2}+2z)}
\eea
For comparison, we recall that the inverse temperature for a SAdS black hole with spherical horizon is given by $\beta=4\pi l^{2}r_{h}/(l^{2}+3r_{h}^{2})$ \cite{hp}. With the definition of $l$ given above, one sees that, in the large $m_{p}r_{h}$ limit, or for small $z$, the two $\beta$'s coincide. On the opposite, in the large $z$ limit we recover the  inverse temperature associated to the Schwarzschild black hole.
For fixed $z$, and similarly to the SAdS case, the temperature diverges for $M\rightarrow 0$ and reaches the absolute minimum $T_{min}=m_{p}/(2\pi \sqrt{2z})\simeq 0.1125\,m_{p}/\sqrt{z}$. For large mass, it grows again and linearly with $r_{h}$.  

As for the SAdS black holes,  the volume of the Euclidean action, obtained by the analytic continuation $t\rightarrow -i\tau$, is infinite. Therefore, the partition function related to the volume $V$ and to the Helmotz free energy $A$ by the relations $\ln Z=-V=-\beta A$ is meaningless. However, along the lines of \cite{hp}, we can define a finite quantity by subtracting the volume of the action with mass $M=0$ to the one with $M\neq 0$ according to the formula
\bea
V=V_{0}\left(\beta_{0}\int_{0}^{\bar r}L(r)dr-\beta\int_{r_{h}}^{\bar r}L(r)dr \right),
\eea
where $V_{0}$ is the volume of the horizon space and $\bar r\gg r_{h}$. The constant $\beta_{0}$ is identified with the arbitrary inverse temperature that corresponds to the $M=0$ background. It is related to $\beta$ by requiring that the periodicity of $g_{tt}$ for the two backgrounds is the same for any large $\bar r$, i.e.
\bea
\beta^{2}_{0}F(r=\bar r,M=0)=\beta^{2}F(r=\bar r,M> 0).
\eea
After some calculations, we find that, in the limit $\bar r\rightarrow\infty$, the normalized volume reads
\bea
V={\pi z x (-2x^{3}+3x+3\arctan(x))\over 3(x^{2}+2)},
\eea
where we have defined the dimensionless variable  $x=m_{p}r_{h}/\sqrt{z}$.   From this result we can compute the energy $E=\partial V/\partial \beta$, the entropy $S=\beta E-V$ and the heat capacity $C=\partial E/ \partial T$. By utilizing the implicit relation between the mass and the horizon radius of the black hole
\bea
{8M\over m_{p}\sqrt{z}}=\arctan(x)+3x\left(1+{x^{2}\over 9}\right),
\eea
we find
\begin{align}
&E=M+{m_{p}\sqrt{z}\,x^{3}(x^{2}+2)^{2}\over 8(x^{2}-2)(x^{2}+1)},\\
&S={\pi z x^{2}(2x^{4}+x^{2}-2)\over (x^{2}+1)(x^{2}-2)},\\
&C= {2\pi z x^{2}(x^{2}+2)(2x^{8}-4x^{6}-11x^{4}-4x^{2}+4)\over (x^{2}+1)^{2}(x^{2}-2)^{3}}.
\end{align}
It is easy to show that all these expressions tend to the values associated to a Schwarzschild black hole when $z$ diverges. We also note that the $z\rightarrow0$ ($z\rightarrow \infty$) limit is the same as the $r_{h}\rightarrow \infty$ ($r_{h}\rightarrow 0$) one, denoting a sort of duality between $z$ and $r_{h}$ (and hence $M$). For large $z$,  the leading term of the entropy is equal to $\pi m_{p}^{2}r_{h}^{2}$ namely a quarter of the horizon area. For $z\rightarrow 0$ the value is just one half. Most importantly, for finite values of $z$ the entropy does not follow the area rule and this is reminiscent of the effects on the entropy induced by high order corrections to the Einstein-Hilbert action, such as in Gauss-Bonnet gravity \cite{anentr}. This fact also reinforces the connection between Galileon theory and higher-dimensional truncated Lovelock theory discussed in \cite{christos}. The energy $E$ and the entropy  $S$ are always positive, except in the interval defined by $0.979\, \sqrt{z}\lesssim m_{p}r_{h}<\sqrt{2z}$ and $0.883\, \sqrt{z}\lesssim m_{p}r_{h}<\sqrt{2z}$ respectively.

\begin{figure}[h!]
\includegraphics[scale=0.44]{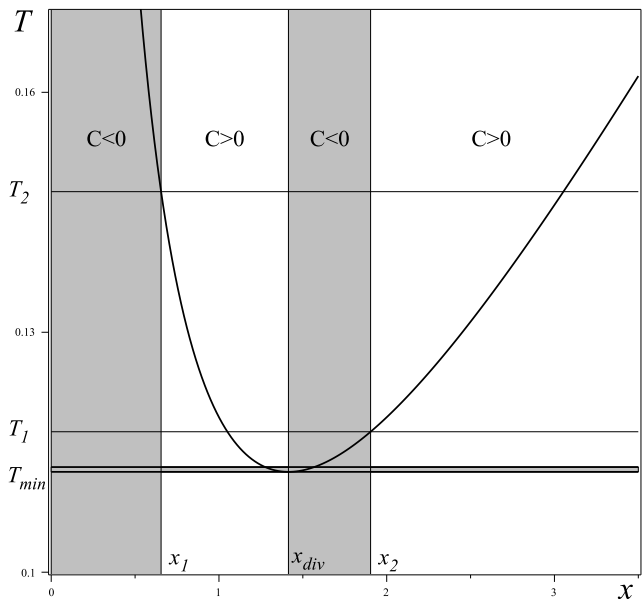}
\caption{\label{plot} Plot of the temperature in units of $m_{p}/\sqrt{z}$ versus $x=m_{p}r_{h}/\sqrt{z}$. In the grey regions the heat capacity is negative. The tiny horizontal strip just above $T_{min}$ represents the region where the free energy $A$ is positive. The relevant quantities are: $T_{min}=0.1125\, m_{p}/\sqrt{z}$, $T_{1}=0.118\,m_{p}/\sqrt{z}$ $T_{2}=0.147\,m_{p}/\sqrt{z}$, $x_{1}=0.655$, $x_{div}=\sqrt{2}$, $x_{2}=1.904$  .}
\end{figure}

The heat capacity $C$ becomes negative for large $z$ and tends to the Schwarzschild value $-2\pi m_{p}^{2}r_{h}^{2}$. Instead $C$ tends to the positive value $4\pi m_{p}^{2}r_{h}^{2}$ for vanishing $z$. As in the SAdS case, it diverges at  the minimum temperature of the black hole $T_{min}$.  The sign of the heat capacity and the behavior of the temperature are shown in Fig.\ (\ref{plot}). We see that for every $T>T_{min}$ there are two black hole solutions with the same temperature. For $T>T_{2}$ the small-$x$ black hole has negative heat capacity, while the large-$x$ one has positive heat capacity. We recall that small $x$ means  small black hole mass or large $z$, that is when the solution tends to the Schwarzschild metric. Thus, for $T>T_{2}$ the situation is the same as for the SAdS case: the small-$x$ black hole decays either into radiation or into the large-$x$ dual \cite{hp}. Between $T_{1}$ and $T_{2}$, however,  there is a region where both black holes are thermodynamically stable as $C>0$, and this marks a difference with respect to the SAdS case. Moreover, between $T_{min}$ and $T_{1}$, the situation is reversed, as the small-$x$ black is stable while the large-$x$ is not.  The Helmholtz free energy $A=\beta^{-1}V$ is positive only in the tiny interval $T_{min}<T<T_{h}$, where $T_{h}\simeq 0.1131\, m_{p}/\sqrt{z}$.

In the SAdS case, for $T_{min}<T<T_{1}$ the black hole with the largest mass is locally stable. However, the evaporation reduces the mass and the temperature until $T_{min}$ is reached and, as the free energy is positive here, the phase transition to pure radiation occurs \cite{hp}. The other black hole solution, with the smallest mass, is unstable in any case as the heat capacity is negative. Therefore, low temperature SAdS black holes are naturally attracted towards the phase transition. In our case the situation is more involved as, in opposition to the SAdS case, the small black hole is stable ($C>0$) in the regime where the temperature increases when $x$ decreases, eventually keeping the system outside the region where $A>0$. It is therefore possible that the phase transition to pure radiation does not occur at this stage. However, as  $x$ keeps decreasing, the black hole temperature eventually  crosses the value $T_{2}$ at which the heat capacity becomes negative. Here, the black hole can dissolve into pure radiation, which collapses into a large $x$ black hole with the same temperature just as in the SAdS case. Therefore, at the end of the the day, the black hole will always end its thermodynamical cycle on the right branch of the temperature curve, where $x$ and $T$ decrease simultaneously, bringing the system towards the usual phase transition at $ T_{min}$.

\section{Other solutions}

To conclude the analysis with $z>0$, we discuss the properties of the scalar field. It is easy to see that $\psi^{2}$ is negative for all $r>r_{h}$, which implies that $\varphi$ is imaginary. The question is whether this implies instabilities in our solution. Examples of  black holes with scalar fields that become complex in some radial range are know in low-energy string theory,  see e.\ g.\ \cite{stringBH}. Generally speaking, in these models the scalar field is complex inside the event horizon, while in our case it is the opposite, so one might worry that the solution is unstable outside the event horizon. Our point of view is that $\varphi$ should not be considered as a matter field but rather as an extra degree of freedom, expressed by the real quantity $\psi^{2}$. This is evident also from the equations of motion (\ref{bern}), where the new degree of freedom appears only as $\psi^{2}$.

For completeness, we briefly comment on the solution with negative $z$. The function  $F$ is basically the same as in Eq.\ (\ref{gtt}) with the $\arctan$ term replaced by arctanh$(m_{p}r/\sqrt{z})$, which reduces the domain of $r$ inside the interval $[0,\sqrt{z}/m_{p}]$. One can find a solution that extends outside this domain by using the formula arctanh$(x)=1/2 \ln[(1+x)/(1-x)]$ and then allowing $1-x\rightarrow |1-x|$, where $x=m_{p}r/\sqrt{z}$ as usual. The result is that,  in the large $r$ limit, the solution tends to a de Sitter space expressed in static coordinates, as expected. However,  the absolute value in $F$ implies that the metric is  non-differentiable at $x=1$, which, in principle, calls for a non-trivial stress tensor to be added to the Lagrangian. We leave to future work a detailed analysis of this solution.

We conclude this paper by discussing the solutions to the system (\ref{bern}) with $K\neq0$.  With a simple argument we can show that there cannot be black hole solutions in this case. The equation of motion  (\ref{current}) can be written as a conservation of a current, $\partial^{r}J_{r}=0$, where
\bea
J_{r}={\psi \sqrt{FG}(m_{p}^{2}r^{2}G+zG-z)\over m_{p}^{2}G^{2}}=-K.
\eea
Now,  a regular horizon located at $r=r_{h}$ implies $F(r_{h})=0$. But then $J^{r}$ must necessarily vanish for all $r$, which means that $K$ must vanish too. Therefore,  black hole solutions are possible only when $K$ vanishes. 
However, there could be other interesting solutions. We first look at Ricci flat solutions, for which $F=G^{-1}=1+Q/r$ for some constant $Q$. Direct substitution into the system (\ref{bern}) shows that the only consistent solution requires again $K=0$. Finally, by using numerical integration, we find a metric that is regular everywhere except at the origin Unfortunately, it turns out that some curvature invariants diverges precisely at $r=0$ so we conclude that when $K\neq 0$ all spherically symmetric solutions host a naked singularity at the origin.
Interestingly, a similar behavior was also observed in the context of intersecting branes in higher-dimensional gravity \cite{rama}.

\section{Conclusions}

In this paper we have explored a brand-new class of black holes that are exact vacuum solutions of scalar-tensor gravity with non-derivative coupling, which is a typical feature of Galileon gravity. They show a locally asymptotic anti-de Sitter geometry and a rich thermodynamical structure, with multiple phase transitions that depend on the mass and on the coupling parameter. The latter turns out to be non-perturbative, in the sense that the usual Schwarzschild solution is recovered when it diverges.

We believe that our results deserve  and need future developments. For example,  the classical stability of the solution should be studied through the analysis of the (coupled) fluctuations of the metric and of the Galileon field. From an astrophysical point of view, it would be interesting to  choose a coupling between matter fields and the Galileon to study the Tolman-Oppenheimer-Volkoff equation and the stability of stars. The solutions that we have found can be also used to test the theory against solar system experiments, in order to constrain the parameter $z$. The asymptotic structure of the solution might be interesting in the context of AdS/CFT correspondence and in brane cosmology, along the lines of \cite{max}. Finally,  black hole solutions should be sought also when more terms belonging to the Galileon family are present in the Lagrangian, such as powers of $\square\varphi$. We hope to report soon on these issues.

We wish to thank R.\ Balbinot, J-P.\ Bruneton, A.\ F\"uzfa, L.\ Heisenberg, L.\ Papantonopoulos and  A.\ Vikman for useful comments. This work is supported by a grant of the ARC 11/15-040 convention.



\begin{thebibliography}{99}

\bibitem{horn}
G.\ W.\ Horndeski, Int.\ J.\ Theor.\ Phys.\ 10, 363 (1974).

\bibitem{fabfour}
C.~Charmousis, E.~J.~Copeland, A.~Padilla and P.~M.~Saffin,
  Phys.\ Rev.\ D {\bf 85} (2012) 104040;
C.~Charmousis, E.~J.~Copeland, A.~Padilla and P.~M.~Saffin,
  Phys.\ Rev.\ Lett.\  {\bf 108} (2012) 051101.
  
\bibitem{galileon}
C.~Deffayet, G.~Esposito-Farese and A.~Vikman,
  Phys.\ Rev.\ D {\bf 79} (2009) 084003;
 C.~Deffayet, S.~Deser and G.~Esposito-Farese,
  Phys.\ Rev.\ D {\bf 80} (2009) 064015.
  
  
\bibitem{massivegrav}
 C.~de Rham, G.~Gabadadze and A.~J.~Tolley,
  Phys.\ Rev.\ Lett.\  {\bf 106} (2011) 231101,
   C.~de Rham and L.~Heisenberg,
  Phys.\ Rev.\ D {\bf 84} (2011) 043503.
  
  
\bibitem{christos} 
  C.~Charmousis, B.~Gouteraux and E.~Kiritsis,
  ``Higher-derivative scalar-vector-tensor theories: black holes, Galileons, singularity cloaking and holography,''
  arXiv:1206.1499 [hep-th].
  
\bibitem{amendola}
L.~Amendola,
  Phys.\ Lett.\ B {\bf 301} (1993) 175.
  
\bibitem{Rcoupling}  
C.~Deffayet, O.~Pujolas, I.~Sawicki and A.~Vikman,
  JCAP {\bf 1010} (2010) 026.
  
  
\bibitem{namur}
  J.~-P.~Bruneton, M.~Rinaldi, A.~Kanfon, A.~Hees, S.~Schlogel and A.~Fuzfa,
  ``Fab Four: When John and George play gravitation and cosmology,''
  arXiv:1203.4446 [gr-qc].
 
\bibitem{Sushkov}
  S.~V.~Sushkov,
  Phys.\ Rev.\ D {\bf 80} (2009) 103505.
  
\bibitem{germani}
C.~Germani and A.~Kehagias,
  Phys.\ Rev.\ Lett.\  {\bf 106} (2011) 161302.
  
 \bibitem{saridakis}
  E.~N.~Saridakis and S.~V.~Sushkov,
  Phys.\ Rev.\ D {\bf 81} (2010) 083510.
  
 \bibitem{greeks}
  T.~Kolyvaris, G.~Koutsoumbas, E.~Papantonopoulos and G.~Siopsis,
  ``Einstein Hair,''
  arXiv:1111.0263 [gr-qc].

\bibitem{nicolis}
L.~Hui and A.~Nicolis,
  ``A no-hair theorem for the galileon,''
  arXiv:1202.1296 [hep-th].

\bibitem{nicolisexp}
L.~Hui and A.~Nicolis,
  ``An observational test of the Vainshtein mechanism,''
  arXiv:1201.1508 [astro-ph.CO].

\bibitem{steer}
C.~Deffayet, X.~Gao, D.~A.~Steer and G.~Zahariade,
   Phys.\ Rev.\ D {\bf 84} (2011) 064039.

\bibitem{damour}
T.~Damour and G.~Esposito-Farese,
  Class.\ Quant.\ Grav.\  {\bf 9} (1992) 2093.
  

\bibitem{danny}
J.~D.~Brown, J.~Creighton and R.~B.~Mann,
  Phys.\ Rev.\ D {\bf 50} (1994) 6394;
D.~Birmingham,
  Class.\ Quant.\ Grav.\  {\bf 16} (1999) 1197.
 
 
  \bibitem{hp}
  S.~W.~Hawking and D.~N.~Page,
  Commun.\ Math.\ Phys.\  {\bf 87} (1983) 577.
  
  \bibitem{peet}
 R.~Kallosh, T.~Ortin and A.~W.~Peet,
  Phys.\ Rev.\ D {\bf 47} (1993) 5400.
  

\bibitem{anentr}
M.~Lu and M.~B.~Wise,
  Phys.\ Rev.\ D {\bf 47} (1993) 3095;
  M.~Visser,
  Phys.\ Rev.\ D {\bf 48} (1993) 583;
R.~-G.~Cai,
  Phys.\ Rev.\ D {\bf 65} (2002) 084014.
  
    
  \bibitem{stringBH}
G.~T.~Horowitz,
  ``The dark side of string theory: Black holes and black strings.,''
  In *Trieste 1992, Proceedings, String theory and quantum gravity '92* 55-99
  [hep-th/9210119];
  R.~Balbinot, L.~Mazzacurati and A.~Fabbri,
  Phys.\ Rev.\ D {\bf 57} (1998) 6185.

  
\bibitem{max}
D.~Birmingham and M.~Rinaldi,
  Mod.\ Phys.\ Lett.\ A {\bf 16} (2001) 1887
  
\bibitem{rama}
S.~K.~Rama,
  ``General Static (Intersecting) Brane Solutions,''
  arXiv:1206.5644 [hep-th];
 S.~K.~Rama,
  ``Static brane-like vacuum solutions in $D \ge 5$ dimensional spacetime with positive ADM mass but no horizon,''
  arXiv:1111.1897 [hep-th].

\end{thebibliography}
\end{document}